\documentclass[12pt]{revtex4}

\usepackage{graphicx}

\begin{document}

\title{Dynamics of Phase Coherence Onset in Bose Condensates of Photons by Incoherent Phonon Emission} 
\author{D.W. Snoke\footnote{email address: snoke@pitt.edu}}

\affiliation{Department of Physics and Astronomy, University of Pittsburgh, Pittsburgh, PA 15260}

\author{S.M. Girvin}

\affiliation{Department of Physics, Yale University, New Haven, CT 06520-8120, USA}

\begin{abstract}
Recent experiments with photons equilibrating inside a dye medium in a cavity have raised the question of whether Bose condensation can occur in a system with only incoherent interaction with phonons in a bath but without particle-particle interaction.  Analytical calculations analogous to those done for a system with particle-particle interactions indicate that a system of bosons interacting only with incoherent phonons can indeed undergo Bose condensation and furthermore can exhibit spontaneous amplification of quantum coherence.  We review the basic theory for these calculations. 
\end{abstract}

\maketitle

\section{Introduction}

Recent experiments \cite{photonBEC} have presented evidence for Bose-Einstein condensation (BEC) and spontaneous development of coherence of photons in a cavity with approximate number conservation of the photons and a photon effective mass. This system is fundamentally similar to the system of polaritons in a microcavity, which has received much attention \cite{pt,deng}. The confinement of the photons to a single longitudinal mode of the cavity gives the photons a dispersion relation which is quadratic in the transverse momentum and hence an effective mass.  The main difference between the two experiments is that in the case of polaritons, the photons are coherently dressed with an exciton component, which gives them an effective hard core particle-particle interaction, while in the photon condensate, the photons are only weakly coupled to the excitations of the medium, so that they are essentially non-interacting.

The photons thermalize almost entirely through emission and absorption of phonons in a dye medium inside the cavity.
The full process is one in which a photon is absorbed by the medium, resulting in an electrical excitation, this electrical excitation incoherently emits or absorbs a phonon, and then a second photon is emitted. Because the emission of the secondary photon has such high likelihood, one can view this entire process as one in which a single photon emits or absorbs a phonon \cite{klaers}.  Under these conditions it is reasonable to expect that the occupancy of different photon modes will equilibrate to the Bose-Einstein distribution with the temperature set by that of the phonon bath.  It is less obvious however that the gas of photons will develop spontaneous phase coherence.  

The observation of BEC behavior of polariton systems via particle-particle interactions, analogous to cold atom condensation, is well known \cite{yama,deveaud,balili,skolnick}. But since in the pure photon system the equilibration occurs almost exclusively via phonon emission and absorption, one can ask to what degree the pure photon system can become a ``true'' Bose condensate \cite{note1}.  
Specifically, since the phonons are part of an incoherent bath, does the interaction with phonons prevent the photon condensate from being phase coherent? The question of obtaining coherent states via dissipative coupling to an incoherent bath has recently also become a topic of interest in the cold atom community, with examples of specific systems tailored to give this result \cite{diehl1,diehl2}.

Our conclusion, based on analytical calculations analogous to earlier calculations \cite{annals} done for a Bose gas with particle-particle interactions, is that a Bose gas at finite temperature can indeed condense and spontaneously amplify phase coherence entirely through incoherent interactions with a bath of phonons. This is consistent with the experimental result of increased coherence seen in interferometry \cite{weitzcoh}. However, unlike the case of interacting polaritons, one expects that the photon condensate will not be superfluid.

The study of the onset of condensation in boson systems has a long history \cite{levich,snokewolfe,kagan,stoof,zoller,svist1}.  It has long been known that a system initially out of equilibrium can quickly build up a peak of particles near zero momentum via particle-particle interactions; Ref.~[\onlinecite{snokewolfe}] found that this occurs in a homogeneous system on the time scale of the classical scattering time. Later, it was shown \cite{moskbook} that bosons equilibrating entirely through interaction with a phonon bath can also build up a zero-momentum peak, as shown in Fig.~1. Both the work showing buildup of the peak in the case of particle-particle interactions and the work showing buildup of the peak in the case of particle-phonon interactions used a quantum Boltzmann equation, which is valid for weakly interacting particles, but which breaks down precisely at the point of long-range coherence in a condensate. The Boltzmann equation method therefore shows that the trajectory of a system is toward condensation, and can give the time scale for the approach to condensation, but does not describe the coherence of the condensate.

The question of the onset of phase coherence has therefore been more debated. There have been two main approaches. In the method of Gardiner and Zoller \cite{zoller}, which has recently been extended to polariton condensates \cite{malp,haug}, all possible Fock states of a finite number of particles in a finite number of states are enumerated, and the transition rates among these states are then calculated. The coherence of the system can then be calculated from the correlation functions computed on an ensemble of different instantiations of the system. The behavior for an infinite system can be taken as the limit of that found for large finite systems. This approach typically involves a somewhat arbitrary cutoff between the quasi-coherent states at low energy and the excited states at higher energy, which are treated with a standard quantum Boltzmann equation.

In the method of classical fields \cite{svist1,svist2,class}, the low-energy states of the system are treated as classical waves, i.e., coherent states, and the behavior of the system is computed using a Gross-Pitaevskii (nonlinear Schr\"odinger) equation; the various coherent classical modes can be viewed as the Fourier components of the general solution of the Gross-Pitaevksii equation. This approach lends itself well to modeling of the spatial variations in the condensate, which relax to long-range order over time. It also implicitly assumes a cutoff between the low-energy classical modes and the excited states modeled by a Boltzmann equation.

A third approach, which we use here, is to write down an equation for the evolution of phase-coherent amplitudes in the system which is analogous to the quantum Boltzmann equation. As for the quantum Boltzmann equation, the validity of this equation will break down exactly at the point of long-range phase coherence, i.e., true condensation, but like the Boltzmann equation, it allows us to see the trajectory of the system toward condensation, and to compute the time scale. It has the advantage of simplicity, and shows quite closely the analogy of phase coherence onset in condensates to other symmetry-breaking systems, e.g. the onset of lasing. It also does not require defining any cutoff between low-energy and high-energy states.

\begin{figure}
\begin{center}
\includegraphics[width=0.5\textwidth]{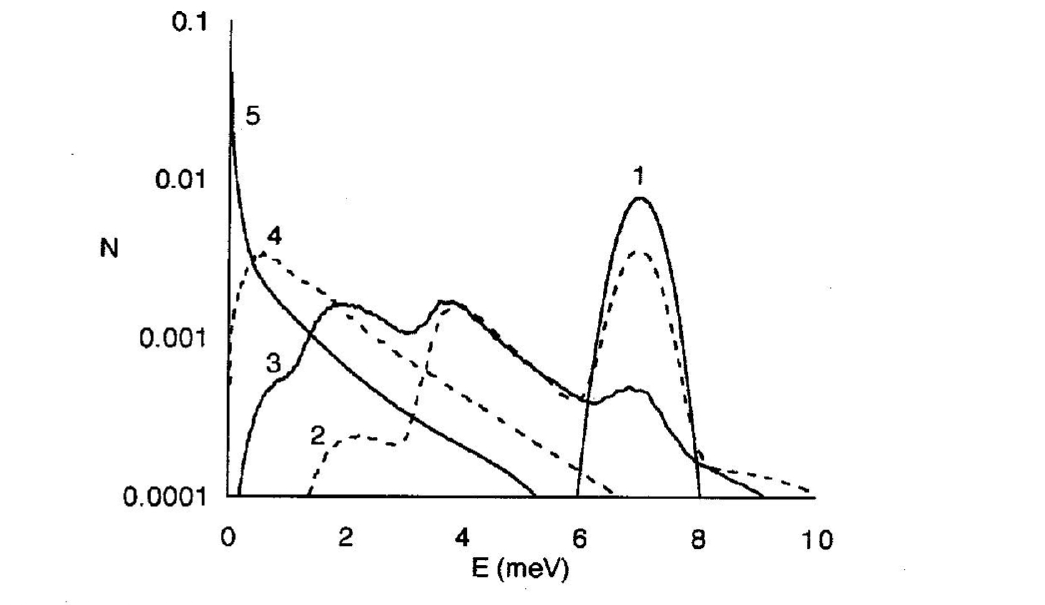}
\caption{The energy distribution of a gas of bosons at various points in time following creation in a non-equilibrium distribution, calculated using a quantum Boltzmann equation which accounts only for interaction with a phonon bath. The phonons are assumed to be in a Planck distribution with constant temperature at all times. From Ref.~[\protect\onlinecite{moskbook}], Section 8.2.}
\label{phonBEC}
\end{center}
\end{figure}
 Ref.~[\onlinecite{annals}] derived the evolution of a closed, many-particle, homogeneous system with weak two-body collisional interactions, with the interaction Hamiltonian
\begin{equation}
{\hat{V}} = \sum_{\{k\}} \hat{V}_{\{k\}}
=\frac{1}{2V}\sum_{{{k}}_1, {{k}}_2, {{k}}_3 } U_{{k}_1, {k}_2,
{k}_3, {k}_4} a^{\dagger}_{{k}_4} a^{\dagger}_{{k}_3}a^{ }_{{k}_2}a^{ }_{{k}_1},
\label{int}
\end{equation}
with $\vec{k}_1+\vec{k}_2 = \vec{k}_3+\vec{k}_4$, and $V$ the volume (we drop vector notation in subscripts for convenience). For a hard-core interaction one assumes $U_{{k}_1, {k}_2,
{k}_3, {k}_4} = U = $ const. The state of the system is the general many-particle state
\begin{equation}
|\psi\rangle = \sum_n \alpha_n |n\rangle,
\label{focksum}
\end{equation}
in which the states $|n\rangle$ are Fock states of the form
\begin{equation}
| n \rangle = \prod_{{{k}} }
\frac{\left(a^{\dagger}_{{k}}\right)^{N_{{k}}}}{\sqrt{N_{{k}}!}}|
\mbox{0}
\rangle,
\label{bath}
\end{equation}
and the $\alpha_n$ are the phase factors for each Fock state.

The rate of change of the expectation value $\langle \hat{N}_k\rangle \equiv \langle \psi| \hat{N}_k|\psi \rangle $, where $\hat{N}_k \equiv a^{\dagger}_ka_k$ is the number operator, can be computing using a quantum Boltzmann equation.
As discussed at length in Ref.~[\onlinecite{annals}], and summarized in the Appendix, writing down a Boltzmann equation relies on the assumption that higher-order, off-diagonal terms of the generalized density matrix, of the form $\langle \hat\rho^{(2)}_{k,k',k'',k'''} \rangle \equiv \langle a^{\dagger}_{k}a^{\dagger}_{k'}a^{ }_{k''}a^{ }_{k'''}\rangle$, are negligible. Taking them as negligible is justified by computing their evolution explicitly.
Ref.~[\onlinecite{annals}] showed that these off-diagonal terms are subject to an evolution equation
 \begin{eqnarray}
&& \frac{d}{dt} \langle \hat{\rho}^{(2)}_{k,k',k'',k'''}\rangle =  \langle\hat{\rho}^{(2)}_{{k,k',k'',k'''}} \rangle \frac{2\pi }{\hbar}\left(\frac{2U}{V}\right)^2\nonumber\\
&&\hspace{.3cm}  \times \frac{1}{2}\sum_{{k}_2,{k}_3}
  \left[\pm
\langle \hat{N}_{k_3}\rangle \langle \hat{N}_{k_4}\rangle (1\pm \langle \hat{N}_{k_2}\rangle) \right.
\nonumber\\
&&\left.\hspace{1cm}
 -\langle \hat{N}_{k_2}\rangle(1\pm \langle \hat{N}_{k_3}\rangle)(1\pm\langle \hat{N}_{k_4}\rangle) \right]\delta(E(\vec{k})) \nonumber\\
&&\hspace{1cm} +\ldots,
\label{secondphase2}
 \end{eqnarray}
where $\delta(E(\vec{k})) = \delta(E_k+E_{k_2}-E_{k_1}-E_{k_4})$ conserves the total energy of the particles in the interaction, and the plus signs are for bosons and the minus signs for fermions. The dots indicate that there are three other terms of the same form, one term for each of the $k$-vectors which appear in $\langle \hat{\rho}^{(2)}_{k,k',k'',k'''}\rangle$.

Note that the right-hand side of (\ref{secondphase2}) is proportional to $\langle \hat{\rho}^{(2)}_{k,k',k'',k'''}\rangle$, which gives this equation the overall form
\begin{equation}
\frac{d}{dt} \langle \hat{\rho}^{(2)}_{k,k',k'',k'''}\rangle = A\langle\hat{\rho}^{(2)}_{{k,k',k'',k'''}}\rangle.
\end{equation}
For fermions, the constant of proportionality $A$ is always negative, but it can be either positive or negative for bosons. When the density is low, so that the average occupation numbers $\langle \hat{N}_k\rangle \ll 1$, the first term in (\ref{secondphase2}), which is due to in-scattering into state $\vec{k}$, is negligible compared to the second, corresponding to out-scattering. Therefore, for fermions at all densities and for bosons in the Maxwell-Boltzmann limit at low density, $A$ is negative and (\ref{secondphase2}) gives dephasing, that is, loss of off-diagonal phase coherence, which in turn justifies the use of the quantum Boltzmann equation which leads to irreversibility. Even in a closed system of only particles interacting with each other by elastic collisions, the large number of degrees of freedom of the excited states act like an incoherent bath.

For bosons at high density, $A$ can be positive, which leads to ``enphasing'' and a breakdown of the quantum Boltzmann equation. There is therefore dramatically different behavior in this case from the dephasing seen in fermion systems and bosons at low density.
The same approach can be used to examine the expectation value of the complex amplitude of the condensate during the process of its growth in non-equilibrium.
The zero and first-order terms give
\begin{eqnarray}
\frac{d \langle a_0 \rangle}{dt} = -i\omega_0  \langle a_0 \rangle - \frac{i}{\hbar}   \frac{UN_{\rm tot}}{V} \langle a_0 \rangle.
\label{condrot}
\end{eqnarray}
In other words, if there is a nonzero coherent amplitude of the condensate, its phase rotates in time with a frequency given by the single-particle energy.
The second term in (\ref{condrot}) is the mean-field renormalization of the energy due to the interactions of the particles, proportional to the total density of particles $N_{\rm tot}/V$.   As discussed in the Supplementary material, there are analogous terms in the time evolution of the $\langle \hat{\rho}^{(2)}_{k,k',k'',k'''}\rangle$ terms for the general case of an interacting gas.

The second-order contribution to the expectation value of the complex amplitude of the condensate gives an equation analogous to (\ref{secondphase2}). As shown in Ref.~[\onlinecite{annals}], this corresponds to the generalized Lindbladian,
\begin{eqnarray}
&&\frac{d}{dt}\langle a_k \rangle = \frac{\pi}{\hbar} \sum_{{\{k\}}{\{k'\}}} \langle \hat{V}_{\{k\}}  a_k\hat{V}_{\{k'\}}-\frac{1}{2}\hat{V}_{\{k\}}\hat{V}_{\{k'\}}  a_k \nonumber\\
&&\hspace{3cm} -\frac{1}{2}a_0\hat{V}_{\{k\}}\hat{V}_{\{k'\}}  \rangle \ \delta(E(\vec{k})),\nonumber\\
\end{eqnarray}
where the $\{k\}$ subscript notation refers to the set of four momentum-conserving $k$-vectors, as defined in (\ref{int}).
After substituting in the definition of the interaction Hamiltonian, one finds
 \begin{eqnarray}
&&\frac{d}{dt} \langle a_k \rangle = \langle a_k \rangle \frac{\pi }{\hbar}\left(\frac{2U}{V}\right)^2
\sum_{{k}_2,{k}_3}
 \left[
\langle \hat{N}_{k_3}\rangle \langle \hat{N}_{k_4}\rangle (1+ \langle \hat{N}_{k_2}\rangle)
\right.  \nonumber\\
&&\hspace{1.5cm}\left.
-\langle \hat{N}_{k_2}\rangle(1+ \langle \hat{N}_{k_3}\rangle)(1+\langle \hat{N}_{k_4}\rangle) \right]   \delta(E(\vec{k})) .\nonumber\\\label{BECphase}
 \end{eqnarray}
The first term in the square brackets gives the total in-scattering rate, and the second term is the total out-scattering rate.  When there is net influx into state $\vec{k}$, any small coherent amplitude will be amplified, with exponential growth of the amplitude of the phase-coherent term. The onset of coherence in a condensate is therefore very much like the onset of coherence in a laser due to spontaneous symmetry breaking (see Ref.~[\onlinecite{snoke-book}], Section 11.3.1, for a review). If the system is perfectly symmetric, it will (ignoring quantum noise)  
never undergo symmetry breaking, but any tiny term that leads to a small coherent amplitude will be exponentially amplified.

The exact same methods can be applied to the case of particles interacting with a phonon bath, through a Hamiltonian term of the form
\begin{eqnarray}
\hat{V} = \frac{1}{\sqrt{V}}\sum_{k,k'} iU_{k'}\left(b^{\dagger}_{k'}a^{\dagger}_{k-k'}a^{ }_k -b^{}_{k'}a^{\dagger}_{k+k'}a^{ }_k\right),
\label{phonint}
\end{eqnarray}
where the $b_{k'}$ operators are the phonon bosonic operators. (For a derivation of this interaction see Ref.~[\onlinecite{snoke-book}], Section 5.1). Of course, a coherent elastic interaction between photons can be derived from this interaction by virtual exchange of a phonon, but this is a higher-order process with two photon-phonon vertices, which means that a virtual-phonon coherent scattering process will give scattering rates proportional to the fourth power of $U_{k'}$ in the quantum Boltzmann equation, as opposed to the second power for single-vertex, incoherent photon-phonon interactions. We can therefore ignore the higher-order coherent elastic processes unless we find that the lower-order, single-vertex incoherent processes give no contribution to the onset of coherence. Experimentally, while there is some evidence for a very weak nonlinear photon-photon interaction in the photon condensation experiments \cite{weitzprivate}, the time scale for thermalization due to these interactions is much longer than the observed thermalization time. In any case, we are interested in the theoretical question of whether coherent elastic interactions are {\em necessary} for a coherent condensate. We find that they are not.

As discussed above, a quantum Boltzmann equation can be derived for the interaction (\ref{phonint}), and this can be evolved using the same numerical approach to give results like those shown in Fig.~1. In the same way, we can derive the evolution of the expectation value of the complex amplitude, to get an equation analogous to (\ref{BECphase}),
following the procedure used in Ref.~[\onlinecite{annals}]. A somewhat tedious but straightforward calculation gives
\begin{eqnarray}
&&\frac{d}{dt} \langle a_0 \rangle = \langle a_0 \rangle \frac{\pi }{\hbar}
\sum_{k'} \frac{U_{k'}^2}{V}
\biggl[\delta(E_{k_0}+E_{k'}^p-E_{k_0+k'})\nonumber\\
&& \hspace{1cm} \times (\pm (1+ N_{k'}^p)N_{k+k'}  - N_{k'}^p(1\pm N_{k_0+k'}))\nonumber\\
\nonumber\\
&&  \hspace{1cm} + \delta(E_{k_0}-E_{k'}^p-E_{k_0-k'})\nonumber \\
&& \hspace{1cm}  \times (\pm N_{k'}^pN_{k_0-k'} -(1+N_{k'}^p)(1\pm N_{k_0-k'}) )\biggr],\nonumber\\
\label{phonphase}
\end{eqnarray}
where $N^p_k = \langle b_k^{\dagger}b^{ }_k\rangle $ is the phonon occupation number and $N_k = \langle a^{\dagger}_ka^{ }_k \rangle $ is the occupation number of the number-conserved particles; similarly, $E^p_k$ is the energy of the phonons and $E_k$ is the energy of the number-conserved particles.
As earlier, the $+$ sign applies when the $a_k$ operators are for bosons and the $-$ sign when they are fermions. A similar result has been found by Laussy and coworkers \cite{malp} for the case of polaritons in microcavities, based on approach of Gardiner and Zoller discussed above. 

This equation has the same underlying physics as (\ref{BECphase}). Two terms are due to in-scattering processes, and two are due to out-scattering; there are four terms because each process can involve either emission or absorption of a phonon. We can therefore conclude that bosons with no mutual interaction can still undergo ``true'' Bose-Einstein condensation, with the associated enphasing, entirely by means of interaction with a phonon bath. This appears to be the case with the photon condensate \cite{photonBEC,weitzcoh}. The incoherence of the phonons does not destroy the coherence of the photons. 

On the other hand, we note that the critical velocity for superfluidity is given by $v = \sqrt{nU/m}$, where $n$ is the particle density, $m$ is the effective mass, and $U$ is the particle-particle interaction which appears in (\ref{int}) (see, e.g. Ref.~[\onlinecite{snoke-book}], Section 11.1.4). Therefore, when the mutual interaction is zero, the critical velocity is zero, and an even slightly flowing condensate is unstable to excitation. A photon condensate with only interaction with a phonon bath is seen experimentally to have coherence \cite{weitzcoh}, which is confirmed by our calculation here, but unless one can formulate a theory of superfluidity by means of the phonon interaction, one would not expect superfluidity of the photons \cite{note2}.    

This points out that quantized vortex formation is not synonymous with superfluidity. The appearance of quantized vortices stems fundamentally from having a single-valued wave function, i.e., phase coherence (see, e.g., Ref.~[\onlinecite{snoke-book}], Section 11.1.4), which as we have seen here, is possible even in a system with a critical velocity of zero. Quantized vortices have been seen in a pure photon system \cite{weiss}, which has many similarities with the photon BEC system of Ref.~[\onlinecite{photonBEC}] except that the condensate was formed by direct pumping rather than by thermally-induced spontaneous symmetry breaking. True superfluidity entails a robustness of the phase coherence of the system against excitation, in other words, a resistance to fracture into nearby coherent states, or conversely, if the system is initially fractured, growth of one coherent mode at the expense of others. This occurs naturally in systems with particle-particle interactions, which give rise to a standard nonlinear Schr\"odinger equation \cite{svist2}. These particle-particle interactions are well established in polariton condensates, as seen, for example, in the mean-field blue shift of the ground state predicted by (\ref{condrot}) and seen in the experiments, and are believed to underly the stability of vortices in polariton condensates \cite{vort1,vort2}.

It is not too hard to see that the basic result of enphasing should apply to any boson system, including ones without number conservation, such as a gas of phonons. Whenever occupation numbers exceed unity, there can be amplification of coherent phase fluctuations. It can be argued that this is the fundamental reason why the low-frequency modes of interacting macroscopic systems such as water waves are always in definite-amplitude states, i.e., act as classical waves instead of Fock states. A macroscopic Fock state, which is a superposition of all different phases, e.g., a water wave which simultaneously has a crest and a trough, is a physically possible state, but is never observed (and it is hard to imagine what it would look like); the reason such is never seen is related to the spontaneous increase of phase coherence in low frequency modes of interacting  
boson systems and the more natural coupling of the environment to the coherent amplitude rather than the boson number.
Conversely, `cat states' of non-interacting photons of even relatively low number decohere quickly \cite{cat} due to dissipation via photon loss.  

The crossover of classical wave behavior to Bose condensate behavior has recently been nicely demonstrated in an optical experiment called ``BEC of classical waves'' \cite{classicalBEC}, in which the time domain of the evolution in the quantum Boltzmann equation was mapped to a spatial coordinate in a nonlinear crystal. This system behaves exactly like the scenario assumed in Ref.~[\onlinecite{svist2}], in which a random ensemble of coherent classical waves evolves toward a single coherent state.
In both this system and the polariton BEC, the mutual interaction of the particles is what leads to thermalization and condensation. In the case of the polariton BEC, essentially what is done is that the nonlinear $\chi^{(3)}$ terms are increased dramatically by putting the photons in resonance with an exciton state \cite{snoke-chapter}. These dressed photons, or polaritons, then can undergo BEC by means of their mutual interaction. The ``photon BEC'' of Weitz and coworkers, by contrast, reaches BEC by interactions with phonons in an incoherent Planckian bath. All three of these experiments are different from standard lasing. None of them require inversion of the medium \cite{snoke-chapter}, but all three obtain coherence and thermalization in a system of particles which can be treated as number-conserved.

We have now before us a range of experiments that can all be described as photonic BEC, with different variations. In one class are those which use photon-photon interactions (i.e., third-order nonlinear wave-mixing terms) to induce BEC in a process analogous to atomic condensates with mutual atom-atom interactions. The electric field plays a role analogous to the matter wave function in cold atom condensates \cite{snoke-chapter}. This class of photonic condensates can be described by a nonlinear Schr\"odinger equation for the condensate, and should have stable superfluidity. Among these there is a range of ratios of particle lifetime to interparticle scattering time. Another class has negligible photon-photon interaction but can still condense and become phase coherent via interaction with an incoherent phonon bath, as shown here. It is an open question whether this type of system can also be superfluid via higher-order interactions, which are extremely weak but could give a very low but nonzero critical velocity.

{\bf Acknowledgements}. This work has been supported by the National Science Foundation through grants DMR-1104383 and DMR-1004406.

\appendix

\section{Derivation of the evolution equations, and numerical solution}

In this section we briefly summarize the method of Ref.~[\onlinecite{annals}] and the numerical methods used to get results like those shown in Fig.~1. For a review of experimental applications of this method, see Ref.~[\onlinecite{boltzreview}].

The change in the average number of particles in state $\vec{k}$ is given by
\begin{eqnarray}
d\langle \hat{N}_{{k}}\rangle &=& \displaystyle \langle \psi_t | \hat{N}_{{k}} | \psi_t \rangle -
\langle \psi_i | \hat{N}_{{k}} | \psi_i\rangle \label{comm}\\
\nonumber\\
&=& \displaystyle\langle \psi_i | e^{(i/\hbar)\int {\hat{V}}(t) dt}  \hat{N}_{{k}} e^{-(i/\hbar)\int {\hat{V}}(t)
dt}| \psi_i \rangle \nonumber\\
&& \hspace{1cm} - \langle \psi_i | \hat{N}_{{k}} | \psi_i\rangle
\nonumber\\
\nonumber\\
&=& \displaystyle\langle \psi_i | e^{(i/\hbar)\int {\hat{V}}(t) dt} {[} \hat{N}_{{k}}, e^{-(i/\hbar)\int
{\hat{V}}(t) dt} {]} | \psi_i \rangle , \nonumber
\end{eqnarray}
where  $\hat{V}(t) = e^{iH_0t/\hbar}\hat{V}e^{-iH_0t/\hbar}$ in the standard interaction notation. The exponential terms are expanded, and only terms up to second order are kept, which is valid for weak interactions.

For the two-body interaction (\ref{int}), the lowest-order term of the expansion is
\begin{eqnarray}
d\langle \hat{N}_{{k}}\rangle &=& \frac{t}{2i\hbar}\sum_{{k}_1, {k}_2}(U_{D} \pm
U_{E})  \left( \langle \psi_i |a^{\dagger}_{k} a^{\dagger}_{k_3}a_{k_2}a_{k_1} | \psi_i \rangle \right. \nonumber\\
&& \left.  -
 \langle \psi_i | a^{\dagger}_{k_1} a^{\dagger}_{k_2}a_{k_3}a_{k} | \psi_i \rangle  \right),
 \label{firstorder}
\end{eqnarray}
where $U_D$ and $U_E$ are the direct and exchange interaction terms, with $+$ for bosons and $-$ for fermions.
If $|\psi_i\rangle$ is a Fock state, this term vanishes exactly; if the two creation operators restore the same two states that the destruction operators removed, then the operator has the form $N_kN_{k_1}$ and is equal for both terms, which cancel. More generally, if $|\psi_i\rangle$ is not a Fock state, this term depends on the value of ``off-diagonal'' elements of the form $\langle \hat{\rho}^{(2)}_{k,k_3,k_2,k_1}\rangle \equiv \langle \psi_i |a^{\dagger}_{k} a^{\dagger}_{k_3}a_{k_2}a_{k_1} | \psi_i \rangle$.

For the second-order term in the expansion, as in the standard derivation of Fermi's golden rule (e.g., Ref.~[\onlinecite{snoke-book}], Section 4.7), when the states are close enough together to be taken as a continuum, one finds terms of the form
\begin{eqnarray}
&&   \left(\int_{0}^t dt' \ e^{i\omega t'}\right)
 \left(\int_0^t dt'' \ e^{-i\omega t''}\right)\nonumber\\
&&\hspace{.5cm}= \
\left|\frac{e^{i\omega t}-1}{\omega}\right|^2 .
\label{fk1}
\end{eqnarray}
which becomes $ \delta(\omega)2\pi t$ in the limit of large $t$; i.e., this term is also linear in $t$. One can therefore pick $t$ to be a small interval $dt$ and divide, to get a rate $d\langle \hat{N}_{{k}}\rangle/dt$. For the two-body interaction (\ref{int}), one obtains the second-order contribution
\begin{eqnarray}
& \displaystyle\frac{d}{dt} \langle \hat{N}_{{k}}\rangle = \displaystyle \frac{2\pi}{\hbar}   \frac{1}{2} \sum_{{k}_1, {k}_2}(U_{D} \pm U_{E})^2  \delta(E_{k_1}\hspace{-.1cm} +E_{k_2} \hspace{-.1cm}- E_{k_3} \hspace{-.1cm}- E_k) \nonumber \\
&\times
\biggl(\langle \psi_i|\hat{N}_{k_1}\hat{N}_{k_2}(1\pm\hat{N}_{k_3})(1\pm \hat{N}_{k})| \psi_i\rangle  \nonumber\\
& - \langle \psi_i|\hat{N}_k\hat{N}_{k_3}(1\pm \hat{N}_{k_2})(1\pm \hat{N}_{k_1})| \psi_i\rangle\biggr)
. \label{qboltz}
\end{eqnarray}
This is almost the standard quantum Boltzmann equation, except that in the standard quantum Boltzmann equation, the averages of products such as $\langle\hat{N}_{k_1}\hat{N}_{k_2}\rangle$ etc.~are factorized into products of averages $\langle\hat{N}_{k_1}\rangle\langle\hat{N}_{k_2}\rangle$; i.e., one assumes no correlations between different $\vec{k}$-states, as in Ref.~[\onlinecite{svist1}]. Ref.~[\onlinecite{annals}] showed that this factorization is valid in the case of no long-range correlations, i.e., before true BEC sets in. The results for phase evolution (\ref{BECphase}) and (\ref{phonphase}) also assume this type of factorization of $\langle a_0\rangle$ and $\langle N_k\rangle $, with the same justification.

In the case of a general state, we must compute the evolution of the ``off-diagonal'' terms which appear in (\ref{firstorder}).
The lowest-order contribution to these factors is
\begin{eqnarray}
 d\langle \hat\rho^{(2)}_{k,k',k'',k'''}\rangle =
  \frac{it}{\hbar}(E_k +\hspace{-.1cm}E_{k'}-\hspace{-.1cm}E_{k''}-\hspace{-.1cm}E_{k'''}) \langle \hat{\rho}^{(2)}_{{k,k',k'',k'''}} \rangle . &&\nonumber\\
  \label{zerorot}
\end{eqnarray}
Thus, if there are nonzero $\hat\rho^{(2)}$ terms, these will rotate in phase with angular frequency proportional to the degree of violation of energy conservation.

The next higher order term is
\begin{eqnarray}
&&\frac{d}{dt} \langle \hat{\rho}^{(2)}_{k,k',k'',k'''}\rangle =
 \frac{2i}{\hbar}U\langle  \left(\hat{N}_{k''} \hat{N}_{k'''}(1\pm \hat{N}_{k})(1\pm \hat{N}_{k'}) \right. \nonumber\\
&&\hspace{2.5cm} \left. -\hat{N}_k \hat{N}_{k'}(1\pm \hat{N}_{k''})(1\pm \hat{N}_{k'''})\rangle\right).
\nonumber\\
\label{firstphase}
\end{eqnarray}
Off-diagonal correlations therefore do in general accumulate due to the interactions. However, the second-order contribution to the evolution of these factors, which is given in Eq.~(\ref{secondphase2}) in the text, suppresses the amplitude of these terms so that they stay small, except in the case of large occupation number of bosons, e.g., when BEC occurs, so that we again have a breakdown of the assumptions of the quantum Boltzmann equation right at the point of condensation.

{\bf Numerical method}. As discussed above, a full evolution equation of the form (\ref{qboltz}) can be simplified by factorizing the averages of products into products of averages. The quantum Boltzmann equations can be reduced to a very tractable form for numerics by the additional assumption of homogeneity in $k$-space. This can be justified either by the assumption of ergodicity (very fast filling of all equal-energy states) or simply by the assumption that the initial state of the system is spatially homogeneous. In this case, the angles of the momentum vectors relative to each other can be integrated over analytically, leaving an integral only over the energies of the particles.

A continuous function $N(E)$ is then defined for the occupation number of the particles, which can deviate from the equilibrium distribution by any amount. This function is represented by a single array of numbers $\{N(E_i)\}$ on a grid of discrete energies $\{E_i\}$, and the rate of change for each energy $E_i$ is calculated using the quantum Boltzmann equation. The value of each $N(E_i)$ is then updated for small $dt$ according to
\begin{equation}
N(E_i(t+dt)) \leftarrow N(E_i)+\frac{dN(E_i(t))}{dt} dt.
\end{equation}
The time step $dt$ is picked to keep the total change of the distribution small during any given time step. With this algorithm, the full evolution of non-equilibrium systems to equilibrium can be determined very efficiently; in single-species systems the numerics can take just a few minutes on a personal computer. These simulations have been fit to data in several experiments \cite{boltzreview}.


\begin{thebibliography}{99}

\bibitem{photonBEC} J. Klaers, J. Schmitt, F. Vewinger, and M. Weitz, Nature {\bf 468}, 545 (2010).

\bibitem{pt} D. Snoke and P. Littlewood, {Physics Today} {\bf 63}, 42 (August, 2010).

\bibitem{deng} H. Deng, H. Haug, and Y. Yamamoto,  Rev. Mod. Phys. {\bf 82}, 1489 (2010).

\bibitem{klaers} Deviations from this model, in which the number of photons is not conserved, have been considered in J. Klaers, J. Schmitt, T. Damm, F. Vewinger, and M. Weitz, Phys. Rev. Lett. {\bf 108}, 160403 (2012).

\bibitem{yama} H. Deng, G. Weihs, C. Santori, J. Bloch, and Y. Yamamoto, 
Science {\bf 298}, 199 (2002).

\bibitem{deveaud} J. Kasprzak et al.,  Nature {\bf 443}, 409 (2006).

\bibitem{balili} R. Balili, V. Hartwell, D.W. Snoke,  L. Pfeiffer and K. West,  Science {\bf 316}, 1007 (2007).

\bibitem{skolnick} A.P.D. Love, D.N. Krizhanovskii, D. M. Whittaker, R. Bouchekioua, D. Sanvitto, S. Al Rizeiqi, R. Bradley, M. S. Skolnick, P. R. Eastham, R. Andr\'e, and Le Si Dang, Phys. Rev. Lett. {\bf 101}, 067404 (2008).

\bibitem{note1} {We ignore the fact that the system is a quasi-2D system of finite size.}

\bibitem{diehl1} S. Diehl, A. Micheli, A. Kantia, B. Kraus, H.P. B\"uchler, and P. Zoller, Nature Phys. {\bf 4}, 878 (2008).

\bibitem{diehl2} S. Diehl, W. Yi, A.J. Daley, and P. Zoller, Phys. Rev. Lett. {\bf 105}, 227001 (2010).

\bibitem{annals} D.W. Snoke, Gangqing Liu, and S.M. Girvin, 
Annals of Physics {\bf 327}, 1825
(2012). 

\bibitem{weitzcoh} J. Klaers, J. Schmitt, T. Damm, F. Vewinger, and M. Weitz, Applied Phys. B {\bf 105}, 17 (2011).

\bibitem{levich} E. Levich and V. Yakhot, Phys. Rev. B {\bf 15}, 243 (1977); J. Phys. A {\bf 11}, 2237 (1978).

\bibitem{snokewolfe} D.W. Snoke and J.P. Wolfe, 
Phys. Rev. B {\bf 39}, 4030 (1989).


\bibitem{kagan} Yu. Kagan, in {\em Bose-Einstein Condensation}, A. Griffin, D.W. Snoke, and S. Stringari, eds. (Cambridge University Press, 1995).

\bibitem{stoof} H.T.C. Stoof, in {\em Bose-Einstein Condensation}, A. Griffin, D.W. Snoke, and S. Stringari, eds. (Cambridge University Press, 1995).

\bibitem{zoller} C.W. Gardiner and P. Zoller, 
Phys. Rev. A {\bf 55}, 2902 (1997); D. Jaksch, C. W. Gardiner, and P. Zoller, 
Phys. Rev. A {\bf 56}, 575 (1997);  C.W. Gardiner and P. Zoller, 
Phys. Rev. A  {\bf 58},  536
(1998).

\bibitem{svist1} Yu. Kagan and B.V. Svistunov, Phys. Rev. Lett. {\bf 79}, 3331 (1997).

\bibitem{moskbook} S.A. Moskalenko and D.W. Snoke, {\em Bose-Einstein Condensation of Excitons and Biexcitons and
Coherent Nonlinear Optics with Excitons}, (Cambridge University Press, 2000).

\bibitem{malp} F.P. Laussy, G. Malpuech, A.V. Kavokin and P. Bigenwald, 
J. Phys., Condens. Matter 16 S3665 (2004);
F. P. Laussy, G. Malpuech, A. Kavokin, and P. Bigenwald, 
Phys. Rev. Lett. {\bf 93}, 016402 (2004).

\bibitem{haug} Nguyen Duy Vy, H. Thien Cao, D. B. Tran Thoai, and H. Haug, 
Phys. Rev. B {\bf 80}, 195306 (2009).


\bibitem{svist2} N.G. Berloff and B.V. Svistunov, Phys. Rev. A {\bf 66}, 013603 (2002).

\bibitem{class} C.N. Weiler, T.W. Neely, D.R. Scherer, A.S. Bradley, M.J. Davis, and  B.P. Anderson,  Nature {\bf 455}, 947 (2008).

\bibitem{snoke-book} D.W. Snoke, {\em Essential Concepts of Solid State Physics}, (Addison-Wesley, 2009).

\bibitem{weitzprivate} M. Weitz, private communication. The effective interaction comes from heating of the dye by the phonon emission from the photons, which then can shift the index of refraction and therefore the cavity photon resonance energy. Because it involves heating, this is a very slow process, with frequency in the kHz range, well below the relevant optical frequencies.


\bibitem{note2} Just as in superconducting pairing, a coherent particle-particle interaction induced by exchange of virtual quantum phonons is less effective at high temperatures where real scattering from thermal phonons occurs.  Here, however, since the particles are bosons, condensation and enphasing can still occur.

\bibitem{weiss} M. Brambilla, F. Battipede, L.A. Lugiato, V. Penna, F. Prati, C. Tamm, and C.O. Weiss, Phys. Rev. A {\bf 43}, 5090 (1991); M. Vaupel and C.O. Weiss, Phys. Rev. A {\bf 51}, 4078 (1995).

\bibitem{vort1} K. G. Lagoudakis et al., Science {\bf 326}, 974 (2009).

\bibitem{vort2} D. Sanvitto et al., Nature Phys. {\bf 6}, 527 (2010).

\bibitem{cat} S. Del\'eglise, I. Dotsenko, C. Sayrin, J. Bernu, M. Brune, J-M. Raimond and
S. Haroche; 
Nature {\bf 455}, 510 (2008).

\bibitem{classicalBEC} Can Sun, Shu Jia, C. Barsi, S. Rica, A. Picozzi and J.W. Fleischer, Nature Physics, (2012).

\bibitem{snoke-chapter} D.W. Snoke, 
in {\em Exciton-Polaritons in Microcavities} (Springer Series in Solid-State Sciences {\bf 172}), D. Sanvitto and V. Timofeev, eds., (Springer, 2012). (arXiv:1205.5756)



\bibitem{deveaud-phase} A. Baas, K. G. Lagoudakis, M. Richard, R. Andr\'e, Le Si Dang, and B. Deveaud-Pl\'edran, 
Phys. Rev. Lett. {\bf 100}, 170401 (2008).

\bibitem{boltzreview} D.W. Snoke, Annalen der Physik {\bf 523}, 87 (2011).




\end{thebibliography}
\end{document}